\documentclass[12pt]{article}
\usepackage[]{alpha}
\usepackage{amsmath,amsfonts,epsfig,cite,graphics}
\usepackage{amssymb}
\usepackage{mathrsfs} 
\usepackage{amsbsy,color}
\usepackage{latexsym}
\usepackage{graphicx}
\usepackage{psfrag}
\usepackage{multirow}
\usepackage{booktabs}
\usepackage{rotating}
\usepackage{url}
\usepackage{dcolumn}
\usepackage[T1]{fontenc}

\newcommand{\be}{\begin{equation}}
\newcommand{\ee}{\end{equation}}
\newcommand{\bea}{\begin{eqnarray}}
\newcommand{\eea}{\end{eqnarray}}
\newcommand{\bes}{\begin{eqnarray}}
\newcommand{\ees}{\end{eqnarray}}
\newcommand{\ba}{\begin{array}}
\newcommand{\ea}{\end{array}}

\newcommand{\eq}[1]{eq.~(\ref{#1})}
\newcommand{\Eq}[1]{Eq.~(\ref{#1})}

\newcommand{\Fig}[1]{Figure~\ref{#1}}
\newcommand{\sect}[1]{Sect.~\ref{#1}}
\newcommand{\Sect}[1]{Section~\ref{#1}}

\newcommand{\tab}[1]{Table~\ref{#1}}
\newcommand{\Tab}[1]{Table~\ref{#1}}

\newcommand{\Ref}[1]{Ref.~\cite{#1}}

\newcommand{\nf}{N_\mathrm{f}}

\newcommand{\rmO}{\mathrm{O}}

\newcommand{\mud}{m_\mathrm{ud}}
\newcommand{\ms}{m_\mathrm{s}}

\newcommand{\fpi}{f_\pi}
\newcommand{\fk}{f_\mathrm{K}}
\newcommand{\fpik}{f_{\pi\mathrm{K}}}
\newcommand{\mk}{m_\mathrm{K}}
\newcommand{\mpi}{m_\pi}

\newcommand{\fm}{\mathrm{fm}}
\newcommand{\mev}{\mathrm{MeV}}
\newcommand{\mq}[1]{m_\mathrm{#1}}

\newcommand{\tr}{\mathrm{tr}\,}

\newcommand{\ZA}{Z_\mathrm{A}}
\newcommand{\ZP}{Z_\mathrm{P}}

\pdfoutput=1   

\begin{document}

\preprintno{DESY 16-162, WUB/16-05}

\title{ Setting the scale for the CLS $2+1$ flavor ensembles}

\author[desy,bnl]{Mattia~Bruno}
\author[wup]{Tomasz~Korzec}
\author[desy]{Stefan~Schaefer}

\address[desy]{John von Neumann Institute for Computing (NIC), DESY \\
    Platanenallee 6, D-15738 Zeuthen, Germany}
\address[bnl]{Physics Department, Brookhaven National Laboratory, Upton, NY 11973, USA}
\address[wup]{Department of Physics, Bergische Universit\"at Wuppertal\\
Gaussstr.~20, D-42119, Wuppertal, Germany}
\begin{abstract}
We present measurements of a combination of the decay constants of the light
pseudoscalar mesons and the gradient flow scale $t_0$, which allow to set the
scale of the  lattices generated by CLS with $2+1$ flavors of
non-perturbatively improved Wilson fermions. Mistunings of the quark masses are
corrected for by measuring the derivatives of observables with respect to the
bare quark masses.
\end{abstract}


\maketitle
\section{Introduction}
A lattice scale is a dimensionful quantity which can be used to form
dimensionless ratios of observables with a well-defined continuum limit. In
principle, its choice is arbitrary, however, the precision to which we can
extract the scale on the lattice and the accuracy, to which its experimental
value is known, will affect the precision of the final results.

Here, we will determine two such scales for the setup chosen in the CLS
simulations with  $\nf=2+1$ flavors of $\rmO(a)$ improved Wilson fermions and
the tree-level L\"uscher--Weisz gauge action, which has been described in
detail in \Ref{Bruno:2014jqa}. The two scales are a combination of the
pseudoscalar decay constants $\fpi$ and $\fk$ as well as $t_0$, the gluonic
dimension two quantity introduced by L\"uscher in \Ref{Luscher:2010iy} using
the Wilson flow. 

Other observables commonly used in scale setting are the mass of the $\Omega^-$
baryon~\cite{Toussaint:2004cj}, the $\Upsilon$-$\Upsilon'$ mass splitting
\cite{Davies:2003ik} or the scale $r_0$~\cite{Sommer:1993ce} defined from the force
between two static quarks. For the latter, like for $t_0$, the physical value
is not known from experiment, but has to be computed on the lattice. Still such
quantities can be very useful as intermediate scales due to their high
statistical accuracy and the fact that their definition
does not include  valence quarks. This makes them also useful in studies which
connect results with different flavor content in the sea. The study presented
here is in some aspects similar to the one by QCDSF~\cite{Bornyakov:2015eaa},
where combinations of hadron masses are used to set the scale in the
determination of $t_0$.

The results of lattice QCD simulations are dimensionless ratios of observables.
Since we restrict ourselves to three flavors, with the two light ones
degenerate, these ratios will differ from those found in Nature. Therefore the
choice of input observable will affect the global normalization of the results.
However, the scale also enters in the definition of the physical quark mass
point and therefore also directly affects the ratios of observables. Because of
the latter,  $\nf=2+1$ flavor results become unique only after specifying the
lattice scale and the observables used to set the quark masses. 
Deviations from uniqueness, however, are very small effects as long as one remains 
in the low energy sector of the theory where decoupling holds~\cite{Bruno:2014ufa}.

The paper is organized as follows: In \Sect{s:2} we first give a brief overview
of the ensembles   as well as the observables we consider.
\Sect{s:3} discusses the issues which occur in the extraction of
the masses and matrix elements in the presence of open boundary conditions
in time, as used in the CLS simulations.  In \Sect{s:4} the method to
correct for mistunings by using mass derivatives of the observables is detailed, before
presenting the results in \Sect{s:5} and concluding.
\section{Setup\label{s:2}}

We want to set the scale for the ensembles generated by the CLS 2+1 effort
which use the tree-level $\rmO(a^2)$ improved L\"uscher-Weisz gauge action and
improved Wilson fermions with a non-perturbative
$c_\mathrm{sw}$~\cite{Bulava:2013cta}. Three values of the coupling have been
employed $\beta=3.4$, 3.55 and 3.7, which correspond roughly to a lattice
spacing of $a=0.085\,\fm$, $0.065\,\fm$ and $0.05\,\fm$, respectively. Data
limited to degenerate quark masses is also available at $\beta=3.46$. An overview can be found in
Tab.~\ref{tab:ens}, with ensemble N203 first described in \Ref{Bali:2016umi}.

\begin{table}[tb]
\begin{center}
\small
\begin{tabular}{ccccllcccc}
\toprule
id &   $\beta$ &  $N_\mathrm{s}$  &  $N_\mathrm{t}$  &  $\kappa_u$ & $\kappa_s$ & $m_\pi$[MeV] &   $m_K$[MeV] &  $m_\pi L$\\
 \midrule
H101 & 3.40 & 32 & 96	& 0.13675962 & 0.13675962	& 420 &420  & 5.8\\
H102 & 3.40 & 32 & 96	& 0.136865 & 0.136549339	& 350 &440  & 4.9\\
H105 & 3.40 & 32 & 96	& 0.136970 & 0.13634079		& 280 &460  & 3.9\\
C101 & 3.40 & 48 & 96	& 0.137030 & 0.136222041	& 220 &470  & 4.7 \\
\midrule                                                   
H400 & 3.46 & 32 & 96 & 0.13688848 & 0.13688848& 420 & 420 & 5.2\\
H401 & 3.46 & 32 & 96 & 0.136725   & 0.136725  & 550 & 550 & 7.3\\
H402 & 3.46 & 32 & 96 & 0.136855   & 0.136855  & 450 & 450 & 5.7\\
\midrule                                                   
H200 & 3.55 & 32 & 96	  & 0.137000 & 0.137000   & 420	&420  & 4.3 	\\
N202 & 3.55 & 48 & 128	& 0.137000 & 0.137000   & 420	&420  & 6.5 	\\
N203 & 3.55 & 48 & 128  & 0.137080 & 0.136840284& 340 & 440 & 5.4   \\
N200 & 3.55 & 48 & 128	& 0.137140 & 0.13672086	& 280&  460  & 4.4 	\\
D200 & 3.55 & 64 & 128	& 0.137200 & 0.136601748& 200 &	480  & 4.2		\\
\midrule		                                               
N300 & 3.70 & 48 & 128	& 0.137000 & 0.137000	 & 420	&420  & 5.1 	\\
J303 & 3.70 & 64 & 192	& 0.137123 & 0.1367546608& 260	&470  & 4.1		\\
\bottomrule
\end{tabular}
\caption{\label{tab:ens}List of the ensembles. In the id, the letter gives the geometry, the 
first digit the coupling and the final two label the quark mass combination.}
\end{center} 
\end{table}

Apart from $\beta=3.46$, these ensembles lie along lines of constant sum of the bare quark masses
$a\mq{q}=(1/\kappa_\mathrm{q}-8)/2$ with degenerate light quarks
$\mq{ud}\equiv\mq{u}=\mq{d}$ and an average quark mass $\mq{sym}$.
$\kappa_\mathrm{q}$ is the standard hopping parameter of the Wilson quark
action~\cite{Wilson:1974sk}. Using the quark mass matrix
$M_\mathrm{q}=\text{diag}(\mq{u},\mq{d},\mq{s})$, we therefore have
\begin{equation}
3\mq{sym}=\mathrm{tr}\,M_\mathrm{q}=2\mq{ud}+\mq{s} = \mathrm{const}\ .
\label{eq:sum}
\end{equation}
This line has been chosen, because it implies a constant $\mathrm{O}(a)$
improved coupling~\cite{Bietenholz:2010jr} 
\begin{equation}
\tilde g_0^2=g_0^2\,(1+\frac{1}{3}b_\mathrm{g}\,a\, \tr M_\mathrm{q})\,,
\label{eq:bg}
\end{equation}
irrespective of the knowledge of the improvement coefficient $b_\mathrm{g}$.

To further specify the chiral trajectories, we have to define a point in the
$(\mud,\ms)$ plane through which it is supposed to pass. To this end, we have
used the dimensionless variables
\begin{equation}
\phi_2=8\,t_0\,m_\pi^2 \qquad\text{and} \qquad
\phi_4=8\,t_0\,(\mk^2+\frac{1}{2}m_\pi^2 ) \label{eq:phi}
\end{equation}
with the requirement that the chiral trajectory intersects the symmetric line
$\mq{ud}=\mq{s}$ at $\phi_4=1.15$. Here, $m_\pi$ and $m_\mathrm{K}$ are the masses
of the pseudoscalars corresponding to the pion and the kaon. The value of $\phi_4=1.15$
comes from a preliminary analysis of the quark mass dependence of $\phi_4$;
only the final analysis can tell in how far this chiral trajectory goes through
the point of physical $\phi_4$ and $\phi_2$.

Three points in this strategy need special attention: First of all, \eq{eq:sum}
does not imply a constant sum of renormalized quark masses, which is already
violated to $\rmO(a)$
\begin{equation}
\tr\,M^\mathrm{R} = Z_\mathrm{m} r_\mathrm{m} [(1+a\bar d_\mathrm{m} \tr
             M_\mathrm{q})\tr M_\mathrm{q}+ad_\mathrm{m} \tr M_\mathrm{q}^2]\,,
\label{eq:imp}
\end{equation}
as worked out in \Ref{impr:nondeg}, whose notation we are using.
Secondly, the tuning in $\phi_4$ is correct only up to a certain degree and it
is at the current stage by no means clear that the thus defined trajectories 
also go through the physical quark mass points. The potential mistuning needs to
be taken into account in the analysis.

Furthermore, the definition of the point of physical quark masses depends at
finite lattice spacing on the scale. The tuning has been done using $t_0$ for
which the precise experimental value is not known. It might therefore be
preferable to use the decay constants also at finite lattice spacing, but this
will necessarily lead to chiral trajectories which are no longer matched to the
same level of accuracy.

\subsection{Observables}
The physical quantity used here to convert the lattice measurements to physical 
units is the combination of the pseudoscalar decay constants of pion $\fpi$ 
and kaon $\fk$, which along with its next to leading order expansion in $\mathrm{SU}(3)$ chiral 
perturbation theory~\cite{Gasser:1984gg} is given by
\begin{equation}
\begin{split}
\fpik&\equiv \frac{2}{3} (f_\mathrm{K} + \frac{1}{2} f_\pi)\\
&\approx f \left [ 1-\frac{7}{6} L_\pi -\frac{4}{3} L_\mathrm{K} -\frac{1}{2} L_\eta + \frac{16 B \tr M}{3 f^2} (L_5 + 3 L_4) \right]\,.
\end{split}
\label{e:fpikch}
\end{equation}
The a priori unknown low energy constants $L_4$ and $L_5$ (defined at the scale $\mu=4 \pi f$)
appear only in the
$\tr M$ term and logarithms are given by $L_x=m_x^2/(4 \pi f)^2 \mathrm{ln}\,
(m_x^2/(4 \pi f)^2)$. A constant  $\tr M$  therefore implies a constant $\fpik$
up to logarithmic corrections.
Note that due to \Eq{eq:imp} we also expect $\rmO(am)$ effects to violate the 
constancy of this combination.

With $\fpik$ as a scale, we can define a second set of dimensionless variables
\begin{equation}
y_\pi = \frac{m_\pi^2}{(4 \pi \fpik)^2} \qquad\text{and} \qquad y_\mathrm{K} =
\frac{m_\mathrm{K}^2}{(4 \pi \fpik)^2} 
\end{equation} 
for which the linear combination $y_{\pi\mathrm{K}}=y_\pi/2+y_\mathrm{K}$ is
again constant in leading order ChPT along our chiral trajectory.

Using the experimental values of the meson masses corrected for isospin breaking 
effects~\cite{Aoki:2016frl} and the PDG values for the decay constants~\cite{Agashe:2014kda},
we use as input parameters
\begin{equation}
\begin{aligned}
m_\pi&= 134.8(3)\mathrm{MeV}\,; & m_\mathrm{K}&= 494.2(3)~\mathrm{MeV}\\
f_\pi&=130.4(2)\,\mathrm{MeV}\ ;&f_\mathrm{K}&=156.2(7)\,\mathrm{MeV}\,.
\label{eq:phys}
\end{aligned}
\end{equation}

The value of  $f_\mathrm{K}$  comes from a direct experiment only up to the 
contribution of the CKM matrix element $V_{\mathrm{u}\mathrm{s}}$, which
ultimately is extracted using theory input. At the current level of accuracy,
the associated uncertainties are acceptable, however, in the future a 
direct measurement from our simulations will be an interesting verification
of this assumption.

\subsection{Finite volume effects}
The finite spatial volume of the lattices can affect the quantities
we are interested in. A detailed study of these effects is planned in the 
future, but a general requirement is that one has to ensure $f_\pi L \gg 1$
and $m_\pi L \gg 1$ for them to be small. 

For the lattices apart from H200 listed in \Tab{tab:ens}, we have $L\geq 2.4$~fm and $m_\pi L > 4$ throughout.
The chiral perturbation theory prediction\cite{Gasser:1986vb,Colangelo:2005gd}
indicates that the systematic effects on $f_\pi$ and $m_\pi$ are below our statistical
uncertainties on the ensembles which enter our analysis, however, in some cases they are not completely negligible. The largest
finite volume effect is on the N200 ensemble, where it amounts to 70\% of the statistical error.
We therefore apply the one-loop finite volume corrections to all data. The remaining 
effect, not accounted for by this correction, should be significantly below the statistical
uncertainty and can therefore be neglected.

The H200 ensemble is excluded from the analysis, because the finite volume effect in the decay constant is too large.
It is at the same physical parameters as the N202 ensemble, but with $L/a=32$ instead of $48$
and is therefore the only lattice with  $L\approx 2\,\mathrm{fm}$.
The measured finite volume effect between the two volumes in the decay constant is $-2.5(1.0)\%$, where ChPT predicts
a $-0.9\%$ correction. While the accuracy of the data is not high enough for a detailed comparison,
the correction beyond the ChPT prediction is to be significantly smaller for the larger volumes  on which we base our computation.

\section{Open boundaries and hadronic observables\label{s:3}}

Three types of fermionic observables are required for the scale setting in this paper:
the masses and decay constants of the pseudoscalar mesons, as well as the PCAC quark
masses. The open boundary conditions of the gauge field configurations
do not pose a fundamental problem in the analysis due to the fact that the 
 transfer matrix is not changed~\cite{Luscher:2011kk,Luscher:2012av}. Still some parts of the analysis have to be
 adapted because of the  broken translational invariance at the boundaries at $x_0=0$ and $x_0=T$.
By construction, the boundary states share the quantum numbers of the vacuum
and, if source or sink of the two-point functions come close to the boundaries,
the whole tower of these states contributes to correlation functions.

As usual,  pseudoscalar masses and decay constants are extracted from
correlation functions of the pseudoscalar density 
$P^{rs}=\bar \psi^r \gamma_5\psi^s$ 
and the improved axial vector current 
$A_\mu=\bar \psi^r \gamma_\mu\gamma_5 \psi^s +a c_\mathrm{A}  \tilde \partial_\mu P^{rs} $ 
with non-perturbatively tuned coefficient $c_\mathrm{A}$~\cite{Bulava:2015bxa}. 
The two-point functions
\begin{equation}
\begin{split}
f_\mathrm{P}^{rs}(x_0,y_0)&=-\frac{a^6}{L^3} \sum_{\vec x,\vec y} \langle P^{rs}(x_0,\vec x)  P^{sr}(y_0,\vec y) \rangle\,,\\
f_\mathrm{A}^{rs}(x_0,y_0)&=-\frac{a^6}{L^3} \sum_{\vec x,\vec y} \langle A_0^{rs}(x_0,\vec x)  P^{sr}(y_0,\vec y) \rangle \,,
\end{split}
\end{equation}
where $r$ and $s$ are flavor indices,
are estimated with stochastic sources located on time slice $y_0$, for which we choose either
$y_0=a$ or $y_0=T-a$. This choice and the general procedure are suggested by the comparison 
of various strategies in \Ref{Bruno:2014lra}.

\subsection{Excited states and boundary effects}
To obtain the vacuum expectation values, we have to define the plateaux regions in
which excited state contributions can be neglected. As in \Ref{Fritzsch:2012wq},
the general strategy to define plateaux is divided in two steps.
First, we perform preliminary fits including the first excited state, where
the fit interval is chosen such that this model describes the data well by using
a $\chi^2$ test.

In the second step, only the function describing the ground state contribution is used, with the fit range given by the
region where the excited state contribution as determined by the first
fit is negligible compared to the statistical errors of the data.

\begin{figure}
\begin{center}
\includegraphics[width=0.75\textwidth]{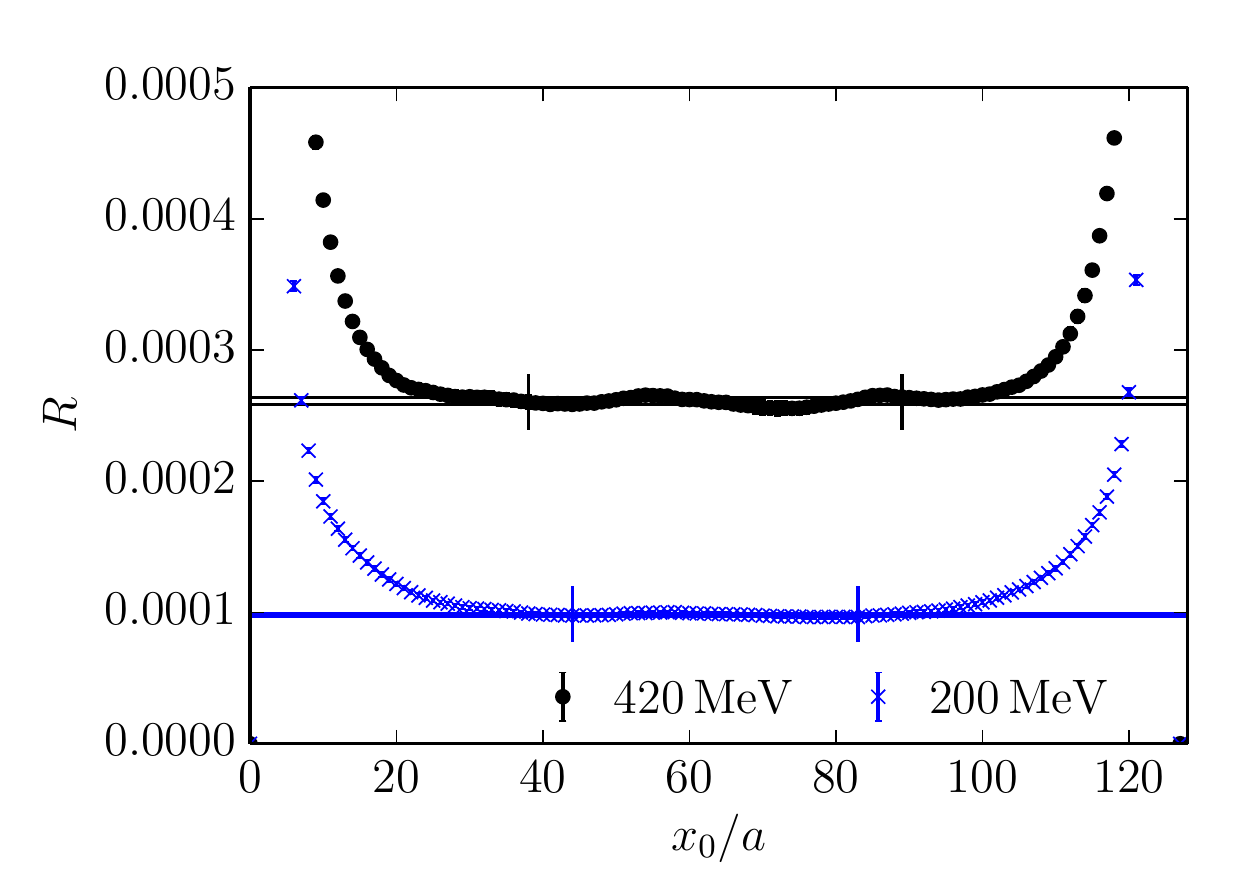}
\end{center}
\caption{\label{f:eff} The effective quantity $R=R_\mathrm{PS}(x_0,a)$ defined
in \Eq{e:R}, from
which the decay constants are extracted, for
ensembles N202 and D200. Both of them share the same $\beta=3.55$, but
differ in the quark masses, the pion having a mass of $\approx 420\,$MeV for the
former and $200$\,MeV for the latter. For the combined observable, the sources are at
$y_0=a$ and $y_0=T-a$, with the sink varying over the temporal extent of the lattice.
The horizontal lines indicate the plateau average and its uncertainty, the vertical lines the plateau ranges. 
Smaller pion mass leads to boundary effects reaching farther into the bulk.}
\end{figure}

From our measurements of $f_\mathrm P$ and $f_\mathrm A$ we observe, at fixed lattice spacing,
that boundary effects increase as the up-down quark masses are lowered. This
turns out to be particularly relevant for the quantity $R_\mathrm{PS}(x_0,y_0)$ defined
below in \Eq{e:R}
where, according to our criterion for the definition of a plateau, boundary
effects can be neglected starting from $x_0 \approx 3~\fm$ for pion masses
around $200~\mev$, as shown in \Fig{f:eff}.   Nevertheless,
despite the fraction of the lattice which is discarded, we have been able to
extract meson decay constants with one percent accuracy (and higher) on all
ensembles, as reported in \tab{tab:raw}. Here, and all other cases presented,
we use the statistical analysis method of \Ref{Schaefer:2010hu}, taking into account 
the autocorrelation of the data including contributions from the slowest observed
modes of the Markov Chain Monte Carlo as determined in \Ref{Bruno:2014jqa}.

For the PCAC masses deviations from a flat behavior constitute a pure
discretization effect. We have observed the largest ones at $\beta=3.4$, where a
plateau can be identified at distances of around $1.7\,\fm$ from the boundaries,
while at $\beta=3.7$ this distance shrinks to $0.7\,\fm$.

\subsection{Meson masses\label{s:mass}}
In the presence of open boundary conditions, the pseudoscalar correlator  $f_\mathrm P$ 
has the asymptotic behavior
\begin{equation} 
f_\mathrm P (x_0,y_0) = A_1(y_0) e^{-m_\mathrm{PS} x_0} +A_2(y_0) e^{-m' x_0} + B_1(y_0) e^{-(E_{2\mathrm{PS}}-m_\mathrm{PS})(T-x_0)} +\dots \,,
\label{eq:mass_sinh}
\end{equation}
for $T\gg x_0\gg y_0$, where we included the contribution from the first excited state.
The third exponential term originates from the first boundary excited state, a finite volume two pion state.
In large volume, $E_{2\mathrm{PS}} \approx 2m_\mathrm{PS}$, leading to the $\sinh$-like functional form 
presented in \Ref{Luscher:2012av}. 

Taking into account the leading corrections from excited states for this formula,
which are exponentially suppressed with the distance of the sink from
the source and the boundaries, respectively, results in
\begin{equation}
am_\mathrm{eff}(x_0)\equiv \log \frac{f_\mathrm{P}(x_0)}{f_\mathrm{P}(x_0+a)}=am_\mathrm{PS} (1+c_1 e^{-E_1 x_0} +c_2 e^{-E_{2\mathrm{PS}} (T-x_0)}+\dots) \,,
\label{e:meff}
\end{equation}
 with $E_1=m'-m_\mathrm{PS}$ and only $c_1$ and $c_2$ depending on the source
position $y_0$. 
As discussed in the previous section, we determine the plateau range in $x_0$,
where the exponential corrections can be safely neglected compared
to the statistical uncertainties. 
The results of the plateau fits can be found in \Tab{tab:raw}.

To check for possible systematics, also direct fits of \Eq{eq:mass_sinh} including
terms of excited states have been tried, without going through the effective mass.
The differences of the results are significantly below our statistical accuracy.

\begin{sidewaystable}
\begin{center}
\footnotesize
\begin{tabular}{cccccccc}
\toprule
id &  $t_0/a^2$ & $am_\pi$ &   $am_\mathrm{K}$ & $am_{12}$ & $am_{13}$ & $af_\pi$ & $af_\mathrm{K}$ \\
 \midrule
 H101 & 2.8469(59) & 0.18302(57) & 0.18302(57) & 0.009202(45)  & 0.009202(45)  & 0.06351(34) & 0.06351(34) \\ 
 & 2.8619(56) & 0.17979(17) & 0.17979(17) & 0.008893(36) & 0.008893(36) & 0.06296(37) & 0.06296(37) \\ 
 H102 & 2.8801(73) & 0.15412(65) & 0.19165(52) & 0.006520(48) & 0.010187(47) & 0.06057(34) & 0.06369(27) \\ 
     & 2.8861(54) & 0.15306(23) & 0.19068(16) & 0.006428(35) & 0.010091(39) & 0.06053(31) & 0.06358(26) \\ 
 H105 & 2.8933(78) & 0.12185(95) & 0.20127(62) & 0.003956(50) & 0.011258(47) & 0.05723(57) & 0.06388(31) \\ 
     & 2.8920(74) & 0.12151(34) & 0.20166(23) & 0.003952(46) & 0.011304(36) & 0.05752(59) & 0.06432(28) \\ 
 C101 & 2.9080(51) & 0.09759(87) & 0.20645(36) & 0.002494(42) & 0.011872(31) & 0.05561(40) & 0.06383(21) \\ 
     & 2.9044(39) & 0.09901(36) & 0.20708(11) & 0.002566(29) & 0.011935(25) & 0.05573(38) & 0.06390(21) \\ 
\midrule
 H400 & 3.635(13)  & 0.16345(66) & 0.16345(66) & 0.008228(36)  & 0.008228(36)  & 0.05690(37) & 0.05690(37) \\ 
  & 3.662(12)  & 0.15896(28) & 0.15896(28) & 0.007786(63) & 0.007786(62) & 0.05631(52) & 0.05631(52) \\
 H402 & 3.558(16)  & 0.17727(62) & 0.17727(62) & 0.009779(44)  & 0.009779(44)  & 0.05887(41) & 0.05887(41) \\
\midrule
 H200 & 5.150(25)  & 0.13717(76) & 0.13717(76) & 0.006856(30)  & 0.006856(30)  & 0.04704(43) & 0.04704(43) \\
 N202 & 5.164(16)  & 0.13407(43) & 0.13407(43) & 0.006855(16)  & 0.006855(16)  & 0.04829(20) & 0.04829(20) \\ 
 & 5.166(15)  & 0.13382(20) & 0.13382(20) & 0.006832(39) & 0.006832(36) & 0.04829(21) & 0.04829(21) \\ 
 N203 & 5.1433(74) & 0.11224(30) & 0.14369(23) & 0.004751(15)  & 0.007902(12)  & 0.04632(17) & 0.04901(14) \\ 
     & 5.1427(80) & 0.11233(16) & 0.14377(11) & 0.004761(26) & 0.007912(26) & 0.04639(18) & 0.04906(14) \\ 
 N200 & 5.1590(76) & 0.09222(34) & 0.15066(24) & 0.003150(11)  & 0.008649(12)  & 0.04422(18) & 0.04911(20) \\ 
     & 5.1600(76) & 0.09197(20) & 0.15053(11) & 0.003137(22) & 0.008636(19) & 0.04432(19) & 0.04915(20) \\ 
 D200 & 5.1802(78) & 0.06502(35) & 0.15644(16) & 0.001536(12)  & 0.009379(11)  & 0.04233(16) & 0.04928(21) \\ 
     & 5.1761(82) & 0.06611(30) & 0.156912(86) & 0.001591(16) & 0.009436(17) & 0.04253(18) & 0.04943(20) \\ 
\midrule
 N300 & 8.576(30)  & 0.10630(34) & 0.10630(34) & 0.0055046(91) & 0.0055046(91) & 0.03790(20) & 0.03790(20) \\ 
 & 8.596(27)  & 0.10376(16) & 0.10376(16) & 0.005237(47) & 0.005237(37) & 0.03770(23) & 0.03770(23) \\ 
 J303 & 8.613(20)  & 0.06514(35) & 0.12015(19) & 0.002053(17)  & 0.007204(33)  & 0.03424(24) & 0.03854(37) \\
     & 8.637(24)  & 0.06259(28) & 0.11879(11) & 0.001884(44) & 0.007027(67) & 0.03399(36) & 0.03845(50) \\
\bottomrule
\end{tabular}
\caption{\label{tab:raw}Values of $t_0$, the pseudoscalar quark masses and decay
constants as well as the PCAC masses. For each ensemble we give the measured values in the rows with the 
labels and in the row below the values after a shift to $\phi_4=1.11$.}
\end{center} 
\end{sidewaystable}

\subsection{Decay constants and quark masses}
The vacuum expectation values needed for the extraction of the decay constants are obtained from 
the plateaux in $x_0$ of the ratio (where we drop the flavor indices $rs$)
\begin{equation}
R_\mathrm{PS} (x_0,y_0)=\left [ \frac{f_\mathrm{A}(x_0,y_0) \, f_\mathrm{A}(x_0,T-y_0)}{f_\mathrm{P}(T-y_0,y_0)} \right ]^{1/2}\,.
\label{e:R}
\end{equation}
This ratio is formed such that matrix elements of operators close to the boundary drop out.
In this case, the plateaux are defined by fitting the ratios $R_\mathrm{PS}$ with 
\begin{equation}
R_\mathrm{PS} (x_0,y_0)= R (1 + c_1(y_0) \cosh[ -E_1 (T/2 - x_0)] ) \,,
\end{equation}
since it is invariant under time reversal transformations.
Once the relevant matrix element is known, the pseudoscalar decay constants are computed from 
\begin{equation}
f_\mathrm{PS} = \ZA ( \tilde g_0) \big[ 1 + \bar b_\mathrm A a \tr M_\mathrm q + \tilde b_\mathrm A a m_{rs} \big] f_\mathrm{PS}^\mathrm{bare} 
\end{equation}
\begin{equation}
 f_\mathrm{PS}^\mathrm{bare} = \sqrt{\frac{2}{m_\mathrm{PS}}}\, R^\mathrm{aver}_\mathrm{PS} \,, 
\end{equation}
where 
$R_\mathrm{PS}^\mathrm{aver}$ is the plateau average of the ratio previously introduced.

The third observable we are interested in is the PCAC quark mass
$m_{rs} (x_0,y_0) = \tilde \partial_{x_0} f_\mathrm A^{rs}(x_0,y_0) / ( 2 f_\mathrm P^{rs}(x_0,y_0) )$ where
$\tilde \partial_{x_0}$ is the symmetric derivative  in time direction.
With the same technique described for the effective mass in \sect{s:mass},
plateaux in $x_0$ are also found for this quantity, 
which is then multiplicatively renormalized (up to $O(a^2)$ corrections) according to
\begin{equation}
m_{rs,\mathrm{R}} = \frac{\ZA}{\ZP} \big[ 1 + (\bar b_\mathrm A - \bar b_\mathrm P) a \tr M_\mathrm q + 
(\tilde b_\mathrm A - \tilde b_\mathrm P) a m_{rs} \big] m_{rs} \,.
\label{eq:mrsR}
\end{equation}
The present knowledge of the improvement coefficients
for the action that we used is 
limited to one-loop\footnote{Non-perturbatively determined values have become available 
while writing up the present analysis~\cite{Korcyl:2016ugy}.}
perturbation theory~\cite{Taniguchi:1998pf}
\begin{equation}
\begin{gathered}
\tilde b_\mathrm A - \tilde b_\mathrm P =  -0.0012 g_0^2 + O(g_0^4) \,, \\
\tilde b_\mathrm A = 1 + 0.0472 g_0^2 + O(g_0^4) \,, \quad \bar b_\mathrm A = \bar b_\mathrm P = O(g_0^4) \,.
\end{gathered}
\label{eq:bA_bP_1loop}
\end{equation}
The finite renormalization factor $\ZA$ has been computed using the
Schr\"odinger Functional~\cite{Bulava:2016ktf} and its  chirally rotated variant~\cite{mattiatomasz}. 
We use the latter result due to its higher statistical
accuracy, measured directly at the simulated values of $g_0$, thus
neglecting terms of order $b_g g_0^4$. The non-perturbative running of the scale-dependent factor $\ZP$ is
not yet completed \cite{Campos:2015fka}. Hence, in the following we will
consider only ratios of renormalized PCAC masses which do not depend on
renormalization factors. 

Starting from the leading matrix element of $f_\mathrm P$, 
a second possibility to obtain the decay constants is based on the PCAC relation.
The former can be obtained from 
a ratio similar to $R_\mathrm{PS}$ where the axial two point functions in the numerator
are simply replaced by their corresponding $f_\mathrm P$. For this quantity, however, we observed
much stronger boundary contaminations and therefore we did not follow this 
strategy to compute the pseudoscalar decay constants.

\section{Mass corrections\label{s:4}}
As we have seen above, it is necessary to control small corrections in the
quark masses from the ones at which the simulations have been performed. This
could be done by reweighting~\cite{Hasenfratz:2008fg,Finkenrath:2013soa}, but
here  we only consider the leading corrections in a Taylor expansion. Since the
required shifts are typically determined from the fit to the data, this has the
advantage that we can include its effect easily in the full data analysis
without need of interpolation between the  measured points of a reweighting.

For a general function  $f(\bar A_1(m),\dots,\bar A_n(m))$ of expectation
values of primary observables $\bar A_i=\langle A_i \rangle $ , the derivative
with respect to a parameter $m$ of the theory reads
\begin{equation}
\frac{d}{dm} f = \sum_i \frac{\partial f}{\partial \bar A_i } 
\left [ \big
\langle \frac{\partial A_i}{\partial m} \big \rangle 
- 
\big \langle 
(A_i-\bar A_i)(\frac{\partial S}{\partial m} -\overline{\frac{\partial S}{\partial m}} )
\big \rangle 
\right ] \label{eq:dm}
\end{equation}
with $S$ denoting the action of the theory. In the analysis we then use
\begin{equation}
f(m') \to f(m)+(m'-m) \frac{d}{dm} f(m)
\end{equation}
neglecting higher order terms.

\subsection{Measurements}
For the measurement of the derivative we therefore need to compute the explicit derivative 
of the observable as well as the one of the action. If $m$ is the bare quark mass,
the derivative of hadronic correlation functions is easily evaluated as in
\begin{equation}
\partial_{m}\tr \big [ \frac{1}{D+m}\, \Gamma\,  \frac{1}{D+m'}\, \Gamma'\,\big]
=-\tr \big[ \frac{1}{(D+m)^2}\, \Gamma\, \frac{1}{D+m'}\, \Gamma' \big]\,.
\end{equation}
The numerical effort is limited: for each propagator a second inversion on the solution is necessary.

The second term in \Eq{eq:dm} contains the derivative of the action
\begin{equation}
-\partial_m \log \det (D+m) = -\tr (D+m)^{-1}\,,
\end{equation}
which can be evaluated using stochastic estimates of the trace.
For our ensembles, we used 16 sources and found the noise introduced by them to be 
significantly inferior to the gauge noise of this observable.

\subsection{Examples}
\begin{figure}
\begin{center}
\includegraphics[width=0.95\textwidth]{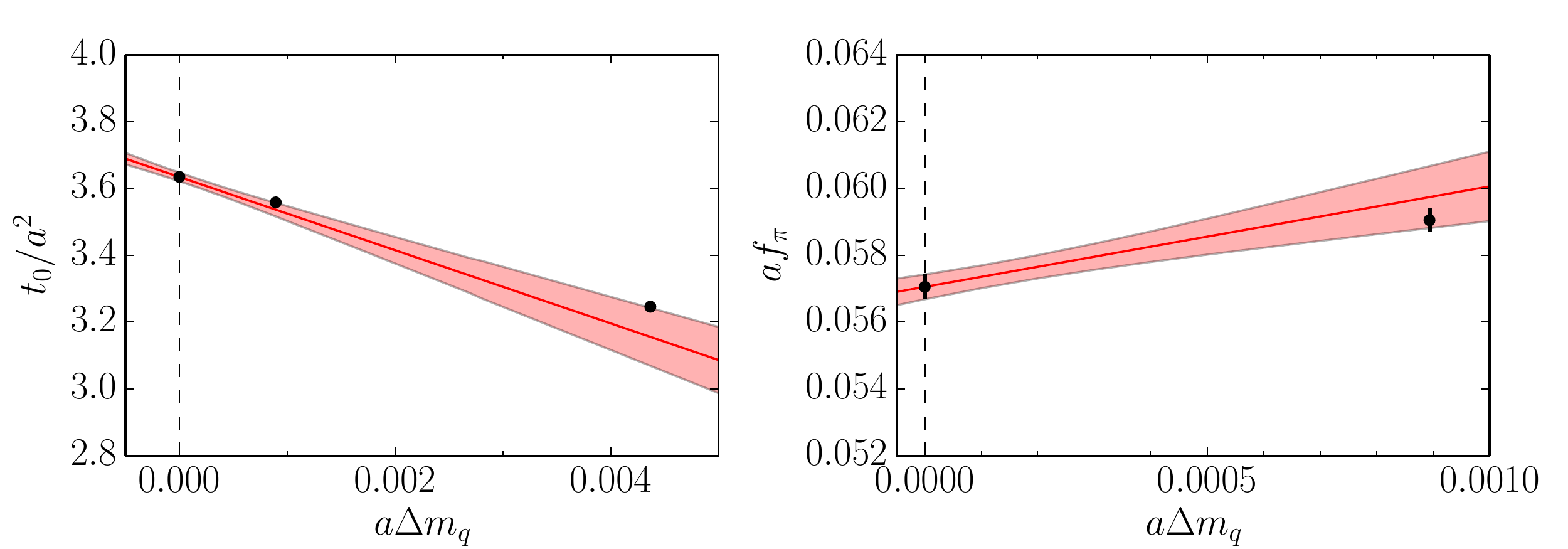}
\end{center}
\caption{\label{f:shift}Examples of the mass shifts for the $t_0$ and $f_\pi$ 
along the symmetric line $m_\mathrm{ud}=m_\mathrm{s}$ at $\beta=3.46$.
The data points correspond to the measurements on the ensembles H400 and
H402 as well as H401 in the case of $t_0$. 
The shaded bands give the linear approximation to the mass dependence starting 
from the leftmost point. 
Deviations
from the linear approximation are smaller than the increase in the statistical uncertainties.}
\end{figure}

To test the method we use ensembles at $\beta=3.46$ which have been generated 
along the symmetric line H400, H401 and H402, where H402 has a sea quark mass
which is roughly 19\% larger than the one of H400 and H401 has roughly twice 
this mass. The results are displayed in \Fig{f:shift}. We give the direct
measurements on the three ensembles as well as the prediction indicated by
the shaded band obtained from ensemble H400.

For $t_0$ we can shift roughly 9\% in the quark mass before doubling the statistical uncertainty,
for the decay constant this level is reached at a 5\% shift.
No deviations from the linear
approximation beyond the statistical error are visible in the displayed region.
Since the shifts we apply in the following are smaller than those, we assume 
that the systematic error from dropping higher orders in the 
Taylor expansion can be neglected compared to the increase in statistical
uncertainty.

\section{Chiral and continuum extrapolation\label{s:5}}

There is no unique choice of chiral trajectory in the $\mud$--$\ms$ plane along 
which one moves as the pion mass is changed, as there is no unique choice
of matching condition between different lattice spacings. These choices, however,
have an impact on the ease with which the extrapolation to physical quark masses
and to the continuum can be performed.

\subsection{Chiral trajectory}
As already mentioned above, the chiral trajectories defined by $\mathrm{tr} M_\mathrm{q}=$\,const
lead to discretization effects of $\rmO(am)$ given in \Eq{eq:imp}. To avoid them,
an improved proxy for the quark mass is needed. There are basically
two options: the PCAC quark masses and the meson masses. The former has
the advantage that a trajectory defined through a constant sum
of these quark masses automatically leads to a constant coefficient of $b_\mathrm{g}$ in
\Eq{eq:bg}.
The disadvantage is that the improvement coefficients $b_\mathrm{A}-b_\mathrm{P}$
and  $\tilde b_\mathrm{A}-\tilde b_\mathrm{P}$ are known only perturbatively. However,
our masses are small and so are the one-loop values of these combinations.

We opt for using the sum of the mass squares of pseudoscalar mesons $\mk^2+\frac{1}{2} \mpi^2$, 
which in leading order of Chiral Perturbation Theory is proportional to the sum
of the three light quark masses. While these do not introduce any discretization 
effects at $\rmO(a)$, it might introduce a small variation of the improved
coupling, since the sum of quark masses varies due to higher order effects in 
ChPT. As we will see below, on a chiral trajectory defined through a constant
$\phi_4=8t_0(\mk^2+\frac{1}{2} \mpi^2)$ also the sum of the renormalized quark 
masses is constant on the per-cent level. We can therefore safely assume that
the effect of a variation in the term coming with $b_\mathrm{g}$ can be neglected.

\subsection{Strategy 1}

The obvious extension of the strategy used in the planning of the simulation is
to continue with $t_0$ as a scale parameter, i.e. finding the physical value of $\phi_4$
along which we move towards the chiral limit. Since the physical value of $t_0$
is not known beforehand, we determine it implicitly from another dimensionful
observable. 

The analysis therefore starts by assuming a certain physical value of $t_0=\tilde t_{0}$.
Together with \Eq{eq:phys}, this defines the target point $( \tilde \phi_2,\tilde \phi_4)$ 
at which we can read off physical results of the calculations.
Starting from the simulated ensembles, shifts along the line 
$\Delta \mq{u}=\Delta \mq{d}=\Delta \mq{s}$ are now performed to reach $\tilde \phi_4$.
This is the direction in which $\mathrm{tr} M_\mathrm{q}$
changes fastest and therefore the effects due to the truncation of the Taylor 
expansion are expected to be smallest at the target  $\tilde \phi_4$.
As an intermediate result we get values of $\sqrt{t_0}/a$,  $a\fpik(\phi_2)$ and their product at constant $\phi_4=\tilde\phi_4$,
which now have to be extrapolated to $\tilde\phi_2$.

For this extrapolation we use two different functional forms, one given by
NLO ChPT, the other a Taylor expansion around the symmetric point. As noted in
\Ref{Bietenholz:2011qq}, along the line adopted in our simulations, the linear
term in the quark mass does not contribute to the Taylor expansion and we can
therefore use $F_\mathrm{T}^\mathrm{cont}(\phi_2)=c_0+c_1(\phi_2-\phi_2^\mathrm{sym})^2$. 

The ChPT
formula $F_\chi(\phi_2)$ can easily be derived from \Eq{e:fpikch}. Note that in NLO
ChPT  $t_0$ is constant along our trajectory at this order\cite{Bar:2013ora}
and we have a straightforward relation between $\phi_2$, $\phi_4$ and the meson masses.
At NLO, the  ratios are therefore unambiguously given by the logarithms predicted by ChPT
\begin{equation}
\frac{\sqrt{t_0} \fpik}{(\sqrt{t_0} \fpik)^\mathrm{sym}}=\frac{\fpik}{ (\fpik)^\mathrm{sym}}
=1-\frac{7}{6}(L_\pi-L_\pi^\mathrm{sym})-\frac{4}{3}(L_\mathrm{K}-L_\mathrm{K}^\mathrm{sym})-\frac{1}{2}(L_\eta-L_\eta^\mathrm{sym})\,.
\label{e:chpt}
\end{equation}

These continuum relations are augmented by a term to account for the leading 
discretization effects.
In general, we adopt
\begin{equation}
\sqrt{t_0} \fpik = F_{\mathrm{T/}\chi}^\mathrm{cont}(\phi_2) + c_{T/\chi} \frac{a^2}{t_0^\mathrm{sym}}\
\label{eq:ans}
\end{equation}
and will see below that our data is well compatible with this ansatz.

One result is a value of  $\sqrt{t_0} \fpik$ at  $\tilde \phi_2$ and $\tilde \phi_4$ 
as defined by $\tilde t_0$. Using the physical value of $\fpik$, this gives a value
of $t_0$ in physical units. The final goal is to find the fixed point, at which this value
agrees with the input $\tilde t_0$. This then defines the physical value of $t_0$ and
in turn the physical value of $\phi_4$. 

\begin{figure}
\begin{center}
\includegraphics[width=0.9\textwidth]{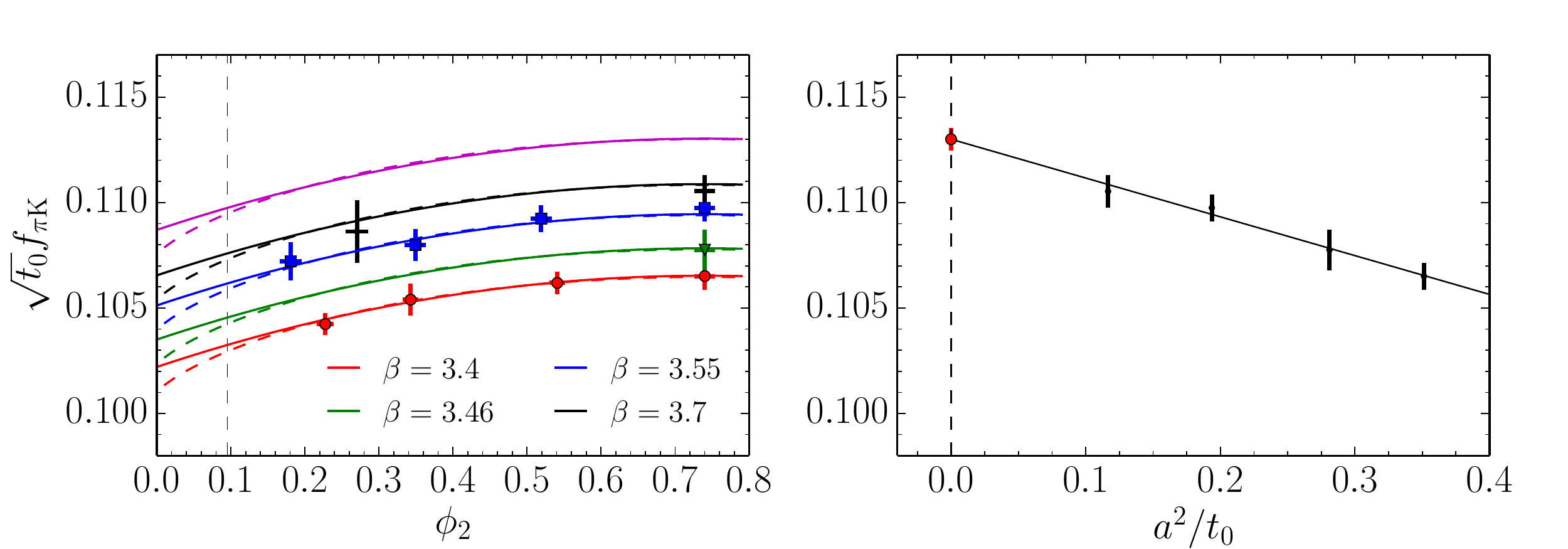}

\includegraphics[width=0.9\textwidth]{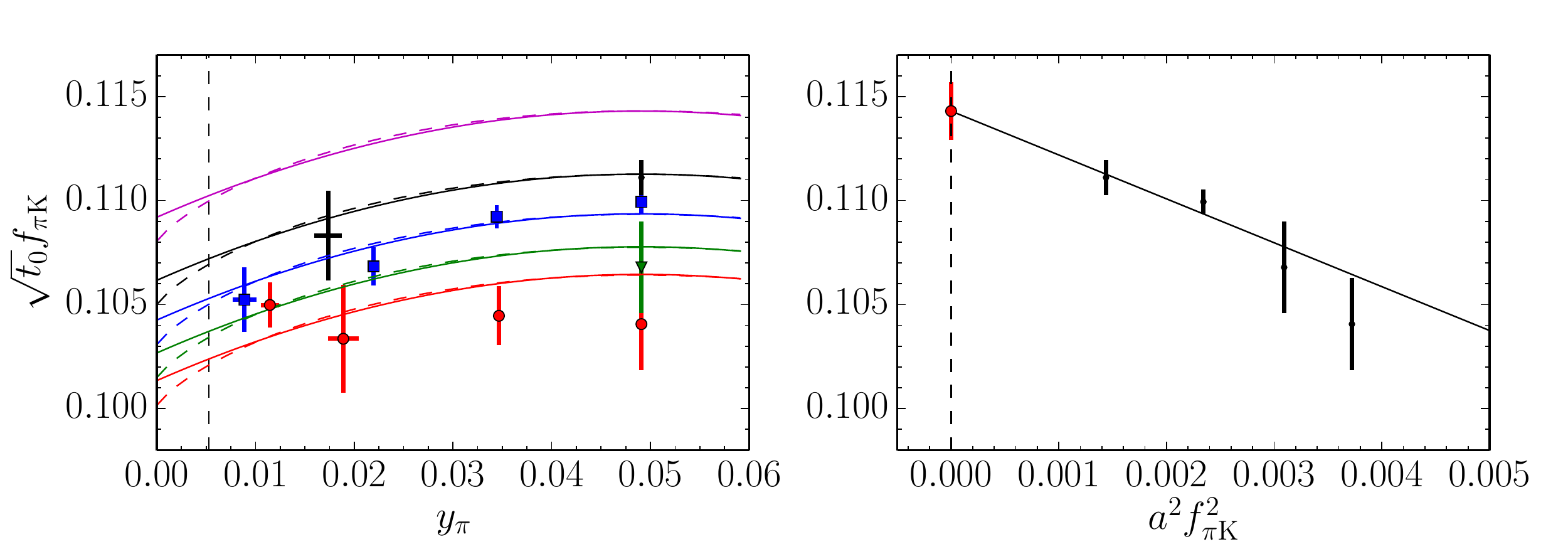}
\end{center}
\caption{\label{f:s1}The dimensionless quantity $\sqrt{t_0} \fpik$ along the line $\phi_4=1.110$ in the top row,
along the line of $y_{\pi\mathrm{K}}=0.074$ in the bottom row. 
In the left panels, we present all measurements as a function of $\phi_2$ together with the fit result
of the quadratic function (solid) and the ChPT \Eq{e:chpt}. The lattice spacing increases from bottom to top,
with the uppermost lines corresponding to the continuum limit. In the right panel the continuum
extrapolation of
the data at the symmetric point is shown.
We observe discretization effects up to $7\%$ for the coarsest lattice spacing.}
\end{figure}

As we see from \Fig{f:s1}, both the ChPT formula as well as the Taylor
expansion,  fitted to our data, hardly differ in the range of our points. Also
at physical quark masses, the difference amounts to roughly half the
statistical uncertainty. However, such a difference might not be enough to
properly quantify the systematic uncertainty associated to the chiral extrapolations. 
More specifically, in ChPT a sensible way to estimate
the size of the higher order terms is by changing the expansion parameter. 
In SU$(2)$ chiral perturbation this is done by using 
either the constant $f$ in the chiral limit 
or $f_\pi(m_\pi)$, which leads to the $x$ and $\xi$ expansions. To mimic this, 
we use either a constant scale proportinal to $\sqrt{t_0}$ or $\fpik(m_\pi)$,
thus leading to $\phi_2$ and $y_\pi$ respectively.

Taking into account the full propagation of the
errors through the fixed point condition, we therefore arrive at physical values of
\begin{equation}
\phi_4^\mathrm{phys}=1.122(16) \qquad \text{and}\qquad \sqrt{8t_0^\mathrm{phys}}=0.4153(29)\,\mathrm{fm}
\end{equation}
for the ChPT ansatz with $y_\pi$. The fit has an excellent quality characterized by a  $\chi^2=8$ 
at 8 degrees of freedom. The quadratic extrapolation gives
\begin{equation}
\phi_4^\mathrm{phys}=1.119(21) \qquad \text{and}\qquad \sqrt{8t_0^\mathrm{phys}}=0.4148(39)\,\mathrm{fm}
\end{equation}
at a  $\chi^2=2.9$, again with 8 degrees of freedom.

By repeating the two fit ansatz with $\phi_2$ as the extrapolation variable, we
obtain a second pair of values for $\phi_4^\mathrm{phys}$: in the case of the Taylor
expansion the difference is negligible,
instead for the ChPT fits the difference is $-0.0022(18)$,
 i.e.  below the statistical accuracy of the final result.
We take this number as our final systematic uncertainty since it covers also the discrepancy
between the results of the Taylor and ChPT extrapolations quoted above.
This leads to
\begin{equation}
\sqrt{8t_0^\mathrm{phys}}=0.415(4)(2)\,\mathrm{fm} \label{e:t01}
\end{equation}
as final result of this strategy. Note that the systematic error also includes possible
uncertainties on the validity range of the chiral extrapolations, which turn out to
be extremely stable (with variations on the 0.5\% level)
under the exclusion, from our fits, of the two most chiral points
and the four symmetric ones.
For convenience, we give the values of our observables shifted to $\phi_4=1.11$
in \Tab{tab:raw}.

With this result we have  fixed the chiral trajectory $\phi_4=\phi_4^\mathrm{phys}$, such 
that now we are in a position to set the scale. 
One method to obtain the lattice spacing in physical units would be to chirally
extrapolate $t_0/a^2$ to $\phi_2^{\rm phys}$ and divide the result by $t_0^{\rm phys}$ of Eq.~(\ref{e:t01}).
We prefer a slightly different method, which avoids this last chiral extrapolation and is instead 
based on directly measured values of $t_0/a^2$ at the symmetric point. \Eq{eq:ans} is fitted to 
$\sqrt{t_0^{\rm sym}}\fpik(\phi_2)$ along the line of  $\phi_4=\phi_4^\mathrm{phys}$.
Dividing the continuum and chirally extrapolated result by the experimental value of $\fpik$ yields
\begin{equation}
   \sqrt{8 t_0^\mathrm{sym}} = 0.413(5)(2)\,\fm\, .
   \label{e:t02}
\end{equation}
The lattice spacings in physical units are obtained by dividing 
$t_0^{\rm sym}/a^2$ by $t_0^{\rm sym}\ [{\rm fm}^2]$
and their values are reported in \Tab{t:lat1}.

\begin{table}[h]
\begin{center}
\begin{tabular}{@{\extracolsep{1cm}}ccc}
\toprule
$\beta$ & $t_0^{\rm sym}/a^2$ & $a [\fm]$   \\
\midrule
3.4  & 2.860(11)(03) & 0.08636(98)(40) \\
3.46 & 3.659(16)(03) & 0.07634(92)(31) \\
3.55 & 5.164(18)(03) & 0.06426(74)(17) \\
3.7  & 8.595(29)(02) & 0.04981(56)(10) \\
\bottomrule
\end{tabular}
\end{center}
\caption{\label{t:lat1}Lattice spacings from strategy 1 set by $t_0$ at the symmetric point and  
physical value of $\phi_4$ as given in \Eq{e:t01}. Note the numbers in the second column are weakly correlated,
whereas the values of the lattice spacings have strong correlations due to \Eq{e:t02}. }
\end{table}

\subsection{Strategy 2}
In the second strategy, we use $\fpik$ to set the scale, shifting each
simulated lattice such that $y_{\pi K}$ equals its physical value 
$y_{\pi K}^\mathrm{phys}=0.07363$. This strategy is simpler since its physical value is
known, see \Eq{eq:phys}. To set the lattice spacing one would compute $a\fpik$
along the line of constant $y_{\pi K}$.

The disadvantage of this approach is that the parameters of our ensembles are farther away from this chiral
trajectory and therefore require larger shifts. This increases the statistical uncertainties and also
potential higher order effects in the Taylor expansion, which we neglect. To show which
accuracy can be reached with the current data, $\sqrt{t_0} \fpik$ is plotted after the shift to physical
$y_{\pi K}$ in the bottom plots of \Fig{f:s1}. As we can see, the statistical uncertainties are significantly larger
than the ones encountered in Strategy 1, such that the applicability of the linear correction terms alone is no 
longer clear. We therefore do not consider this strategy to be competitive on the current data.

Employing the same analysis strategy as in the previous section, using a polynomial function
and the one given by ChPT, we arrive at $\sqrt{8t_0^\mathrm{phys}}=0.417(9)\,\fm$ for the former and
$\sqrt{8t_0^\mathrm{phys}}=0.416(10)\,\fm$ for the latter. 

The advantage of the strategy for the scale setting is a direct value of the lattice spacing from
$\fpik=147.6(5)\,\mathrm{MeV}$ at the physical point. This leads to $a=0.0790(11)\,\mathrm{fm}$,
$0.071(2)~\mathrm{fm}$, $0.0613(9)\,\mathrm{fm}$ and $0.0481(8)\,\mathrm{fm}$ for $\beta=3.4$,
$3.46$, $3.55$ and $3.7$, respectively. The difference to the results in the previous section 
is a discretization effect, which is already visible in the plots on the right hand side of \Fig{f:s1}.

\subsection{Discretization effects}

Because of the large statistical error encountered in strategy 2, we will now restrict ourselves
to the data obtained with the first strategy.
One  assumption entering the analysis presented above is that the
data presented here can be described by the leading discretization effects of
order $a^2$ at the level of statistical accuracy. To get a handle on this, in \Fig{f:s1} the
dimensionless product $\sqrt{t_0}\fpik$ is displayed as a function of $a^2/t_0$
at the symmetric point given by $\phi_4=\phi_4^\mathrm{phys}$. As we can see,
the data exhibits no deviation from a linear behavior, supporting further the
assumption made in the ansatz~\ref{eq:ans}.

\subsection{Chiral extrapolation}

The effect of the chiral extrapolation is best studied by forming ratios between the value
of the observable at the symmetric point and the one at parameters closer to the chiral limit but
at the same lattice spacing. In these ratios, some of the lattice systematics cancels such that for the
chiral effects a high sensitivity can be reached.

\begin{figure}
\begin{center}
\includegraphics[width=0.95\textwidth]{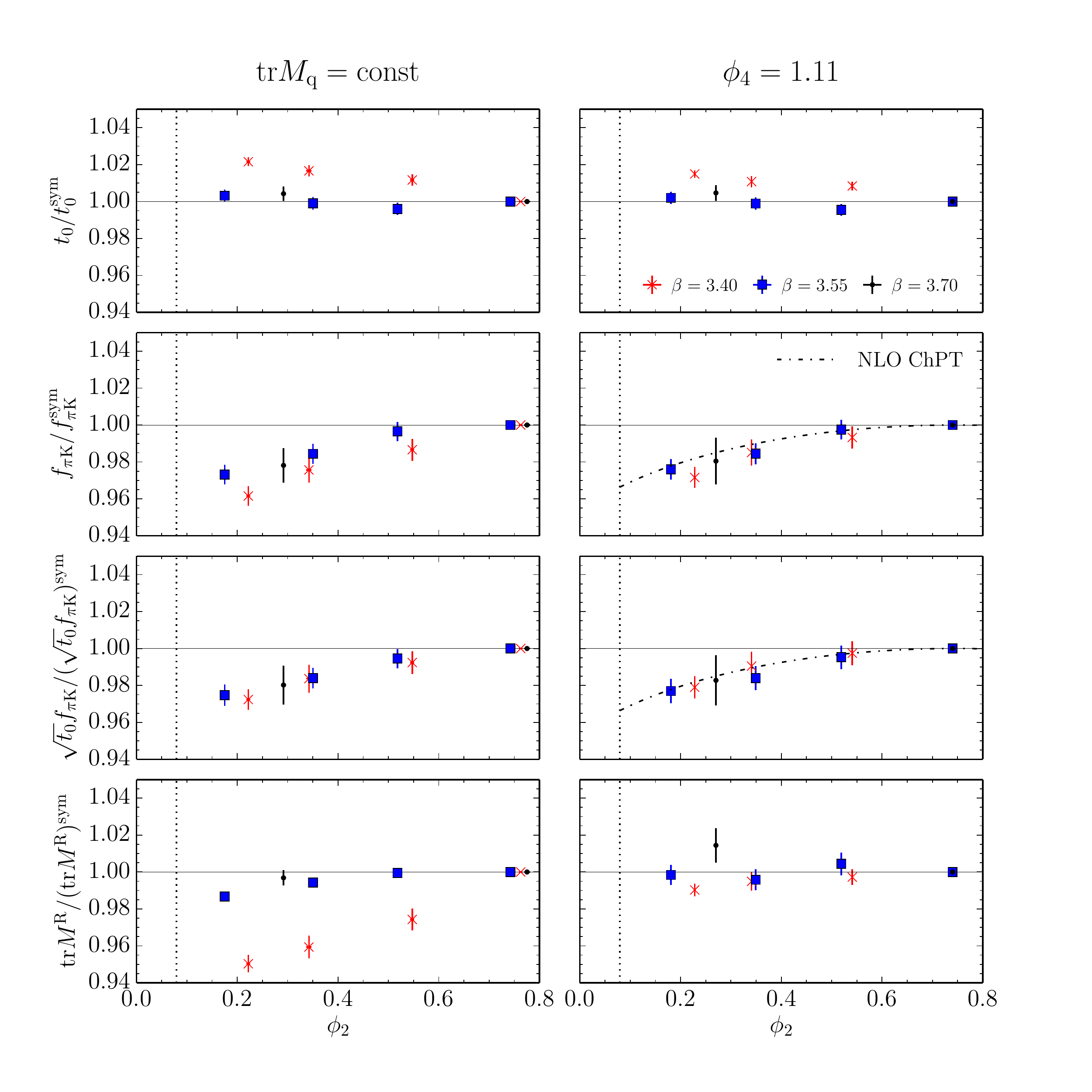}
\end{center}

\caption{\label{f:rat}Effect of the chiral extrapolation on $t_0$, $\fpik$ and
the sum of perturbatively improved PCAC masses. The data is always normalized
by the symmetric point. In the left column, the data as measured on the
simulated ensembles, where $\tr(M_\mathrm{q})$ is kept fixed. On the right
after the shifts to a constant $\phi_4=1.11$ has been applied. In particular
for the quark mass sum we observe a significant effect. While the violation of
$\tr(M_\mathrm{q}^\mathrm{R})$=const before the shift cannot be explained by
effects linear in the lattice spacing, we observe that constant $\phi_4$
implies constant renormalized quark mass to high accuracy.
}

\end{figure}

The original ensembles are along trajectories of constant sum of bare quark masses, matched at the 1\%
level using $\phi_4$ at the symmetric point. The results for the ratios can be found in the left
column of \Fig{f:rat}. As we can see from the lower plot, the sum of renormalized quark masses
is not constant. These masses have been improved with a non-perturbatively determined $c_\mathrm{A}$, effects
of the $b$-terms have been neglected. The fact that the renormalized sum is not constant is a discretization
effect. At $\beta=3.4$ their size is so large that  they cannot be attributed to contributions 
linear in the quark masses alone; higher order contributions are noticeable at this coarse lattice spacing.

In the right column of this plot we see the effect of the shift to a constant $\phi_4=1.11$, which
is close to the physical value. The renormalized quark mass is now constant on the per-cent level even
for the coarsest lattice spacing, with the remaining effects compatible with reasonable values of the $b$-terms.
This also justifies our choice of aiming for a constant $\phi_4$ versus a constant $\mathrm{tr}M^\mathrm{R}$:
the difference between these two options cannot be resolved by the statistical accuracy of the data and in
any case is limited to the per-cent level.

The effect on the ratios of $t_0$ and $\fpik$ is less dramatic. In agreement with the expectation both
based on the Taylor expansion of a flavor symmetric quantity around the symmetric point~\cite{Bietenholz:2011qq,Bornyakov:2015eaa}
and ChPT~\cite{Bar:2013ora}, the chiral corrections are tiny, in particular for the finer lattices.
At the coarsest lattice spacing, some deviation from the constant behavior is still observed, which
is reduced by the shift to $\phi_4=$const.

The chiral effect in $\fpik$ is more noticeable, with a correction on the level
of $3-4\%$ to the physical light quark mass point. Notice, that our data
agrees well with the logarithms predicted by ChPT\footnote{Note that 
in \Eq{e:chpt} we have expanded the denominator to NLO in ChPT. Keeping
the full expression amounts to higher order effects and produces approximately 
a 1\% shift in the ratio at physical quark masses, which is well captured by
the statistical uncertainty of the data in the extrapolations of $\sqrt{t_0} \fpik$
or $a \fpik$.
} in \Eq{e:fpikch}.

\section{Conclusions}
Many observables can be used as a lattice scale, all agree up to effects
which come from an incomplete description of Nature. Here we neglect for 
instance quarks heavier than the strange, electromagnetism and isospin breaking.
The main strategy pursued in the present study is to use $t_0$ as an intermediate
scale, with the approach to the chiral limit along lines of constant $\phi_4=8 t_0 (m_\mathrm{K}^2+m_\pi^2/2)$.
Using $\fpik$ as physical input, this allows the determination of the physical value
of $t_0$. 
This strategy is preferred due to the currently available ensembles in the CLS effort,
because the ensembles have been tuned with $t_0$ as a scale.

Starting with statistical accuracies for the decay constants on the level of $0.5\%$,
we are able to determine $t_0$ at the per-cent level
\begin{equation}
\sqrt{8t_0}=0.415(4)(2)\,\fm\,.
\end{equation}

This compares well to previous determinations using 2+1 flavors by the BMW
collaboration~\cite{Borsanyi:2012zs} that quotes
$\sqrt{8t_0}=0.414(7)\, \mathrm{fm}$ and is also within $2\sigma$ of the QCDSF result~\cite{Bornyakov:2015eaa}
$0.427(7)\, \mathrm{fm}$ as well as
RBC-UKQCD's  value~\cite{Blum:2014tka}  of $0.407(2)\, \mathrm{fm}$. Using 2+1+1
dynamical flavors the MILC~\cite{Bazavov:2015yea} and HPQCD~\cite{Dowdall:2013rya} collaborations find
$\sqrt{8t_0}=0.4005\binom{22}{11}\, \mathrm{fm}$ and $0.4016(22) \,\fm$, respectively, which might be an effect of
the number of flavors in the sea as is the two-flavor result
$\sqrt{8t_0}=0.434(3)\, \mathrm{fm}$~\cite{Bruno:2013gha}.

With additional ensembles becoming available, the analysis presented here will improve.
However, even the accuracies of the current study will already allow to reach
a good precision in many physics projects.

\acknowledgement

We are grateful to our CLS colleagues for sharing the gauge field configurations
on which this work is based. We would like to thank Rainer Sommer for 
continuous encouragement and many useful discussions. 

We acknowledge PRACE for awarding us access to resource FERMI based in Italy at
CINECA, Bologna and to resource SuperMUC based in Germany at LRZ, Munich.
Furthermore, this work was supported by a grant from the Swiss National
Supercomputing Centre (CSCS) under project ID s384. We  are grateful for the
support received by the computer centers.

The authors gratefully acknowledge the Gauss Centre for Supercomputing (GCS)
for providing computing time through the John von Neumann Institute for
Computing (NIC) on the GCS share of the supercomputer JUQUEEN at J\"ulich
Supercomputing Centre (JSC). GCS is the alliance of the three national
supercomputing centres HLRS (Universit\"at Stuttgart), JSC (Forschungszentrum
J\"ulich), and LRZ (Bayerische Akademie der Wissenschaften), funded by the
German Federal Ministry of Education and Research (BMBF) and the German State
Ministries for Research of Baden-W\"urttemberg (MWK), Bayern (StMWFK) and
Nordrhein-Westfalen (MIWF).

This work was supported by the United States Department of Energy under Grant No. DE-SC0012704.

\usebiblio{latt,ALPHA}


\providecommand{\href}[2]{#2}\begingroup\raggedright\begin{thebibliography}{10}

\bibitem{Bruno:2014jqa}
M.~Bruno et~al., \emph{{Simulation of QCD with $N_{f} = 2$ + 1 flavors of
  non-perturbatively improved Wilson fermions}},
  \href{http://dx.doi.org/10.1007/JHEP02(2015)043}{\emph{JHEP} {\bf 02} (2015)
  043}, [\href{http://arxiv.org/abs/1411.3982}{{\tt 1411.3982}}].

\bibitem{Luscher:2010iy}
M.~{L\"uscher}, \emph{{Properties and uses of the Wilson flow in lattice QCD}},
  \href{http://dx.doi.org/10.1007/JHEP08(2010)071}{\emph{JHEP} {\bf 1008}
  (2010) 071}, [\href{http://arxiv.org/abs/1006.4518}{{\tt 1006.4518}}].

\bibitem{Toussaint:2004cj}
D.~Toussaint and C.~T.~H. Davies, \emph{{The Omega- and the strange quark
  mass}},
  \href{http://dx.doi.org/10.1016/j.nuclphysbps.2004.11.129}{\emph{Nucl. Phys.
  Proc. Suppl.} {\bf 140} (2005) 234--236},
  [\href{http://arxiv.org/abs/hep-lat/0409129}{{\tt hep-lat/0409129}}].

\bibitem{Davies:2003ik}
{\scshape Fermilab Lattice, HPQCD, UKQCD, MILC} collaboration, C.~T.~H. Davies
  et~al., \emph{{High precision lattice QCD confronts experiment}},
  \href{http://dx.doi.org/10.1103/PhysRevLett.92.022001}{\emph{Phys. Rev.
  Lett.} {\bf 92} (2004) 022001},
  [\href{http://arxiv.org/abs/hep-lat/0304004}{{\tt hep-lat/0304004}}].

\bibitem{Sommer:1993ce}
R.~Sommer, \emph{{A New way to set the energy scale in lattice gauge theories
  and its applications to the static force and alpha-s in SU(2) Yang-Mills
  theory}},
  \href{http://dx.doi.org/10.1016/0550-3213(94)90473-1}{\emph{Nucl.Phys.} {\bf
  B411} (1994) 839--854}, [\href{http://arxiv.org/abs/hep-lat/9310022}{{\tt
  hep-lat/9310022}}].

\bibitem{Bornyakov:2015eaa}
V.~G. Bornyakov et~al., \emph{{Wilson flow and scale setting from lattice
  QCD}},  \href{http://arxiv.org/abs/1508.05916}{{\tt 1508.05916}}.

\bibitem{Bruno:2014ufa}
{\scshape ALPHA} collaboration, M.~Bruno, J.~Finkenrath, F.~Knechtli, B.~Leder
  and R.~Sommer, \emph{{Effects of Heavy Sea Quarks at Low Energies}},
  \href{http://dx.doi.org/10.1103/PhysRevLett.114.102001}{\emph{Phys. Rev.
  Lett.} {\bf 114} (2015) 102001}, [\href{http://arxiv.org/abs/1410.8374}{{\tt
  1410.8374}}].

\bibitem{Bulava:2013cta}
J.~Bulava and S.~Schaefer, \emph{{Improvement of $N_f=3$ lattice QCD with
  Wilson fermions and tree-level improved gauge action}},
  \href{http://dx.doi.org/10.1016/j.nuclphysb.2013.05.019}{\emph{Nucl.Phys.}
  {\bf B874} (2013) 188--197}, [\href{http://arxiv.org/abs/1304.7093}{{\tt
  1304.7093}}].

\bibitem{Bali:2016umi}
G.~S. Bali, E.~E. Scholz, J.~Simeth and W.~{S\"oldner}, \emph{{Lattice
  simulations with $N_f=2+1$ improved Wilson fermions at a fixed strange quark
  mass}},  \href{http://arxiv.org/abs/1606.09039}{{\tt 1606.09039}}.

\bibitem{Wilson:1974sk}
K.~G. Wilson, \emph{{Confinement of Quarks}},
  \href{http://dx.doi.org/10.1103/PhysRevD.10.2445}{\emph{Phys.Rev.} {\bf D10}
  (1974) 2445--2459}.

\bibitem{Bietenholz:2010jr}
W.~Bietenholz et~al., \emph{{Tuning the strange quark mass in lattice
  simulations}},
  \href{http://dx.doi.org/10.1016/j.physletb.2010.05.067}{\emph{Phys.Lett.}
  {\bf B690} (2010) 436--441}, [\href{http://arxiv.org/abs/1003.1114}{{\tt
  1003.1114}}].

\bibitem{impr:nondeg}
T.~Bhattacharya, R.~Gupta, W.~Lee, S.~R. Sharpe and J.~M. Wu, \emph{{Improved
  bilinears in lattice QCD with non-degenerate quarks}},
  \href{http://dx.doi.org/10.1103/PhysRevD.73.034504}{\emph{Phys.Rev.} {\bf
  D73} (2006) 034504}, [\href{http://arxiv.org/abs/hep-lat/0511014}{{\tt
  hep-lat/0511014}}].

\bibitem{Gasser:1984gg}
J.~Gasser and H.~Leutwyler, \emph{{Chiral Perturbation Theory: Expansions in
  the Mass of the Strange Quark}},
  \href{http://dx.doi.org/10.1016/0550-3213(85)90492-4}{\emph{Nucl. Phys.} {\bf
  B250} (1985) 465--516}.

\bibitem{Aoki:2016frl}
S.~Aoki et~al., \emph{{Review of lattice results concerning low-energy particle
  physics}},  \href{http://arxiv.org/abs/1607.00299}{{\tt 1607.00299}}.

\bibitem{Agashe:2014kda}
{\scshape Particle Data Group} collaboration, K.~A. Olive et~al., \emph{{Review
  of Particle Physics}},
  \href{http://dx.doi.org/10.1088/1674-1137/38/9/090001}{\emph{Chin. Phys.}
  {\bf C38} (2014) 090001}.

\bibitem{Gasser:1986vb}
J.~Gasser and H.~Leutwyler, \emph{{Light Quarks at Low Temperatures}},
  \href{http://dx.doi.org/10.1016/0370-2693(87)90492-8}{\emph{Phys. Lett.} {\bf
  B184} (1987) 83}.

\bibitem{Colangelo:2005gd}
G.~Colangelo, S.~{D\"urr} and C.~Haefeli, \emph{{Finite volume effects for
  meson masses and decay constants}},
  \href{http://dx.doi.org/10.1016/j.nuclphysb.2005.05.015}{\emph{Nucl. Phys.}
  {\bf B721} (2005) 136--174},
  [\href{http://arxiv.org/abs/hep-lat/0503014}{{\tt hep-lat/0503014}}].

\bibitem{Luscher:2011kk}
M.~L{\"u}scher and S.~Schaefer, \emph{{Lattice QCD without topology barriers}},
  \href{http://dx.doi.org/10.1007/JHEP07(2011)036}{\emph{JHEP} {\bf 1107}
  (2011) 036}, [\href{http://arxiv.org/abs/1105.4749}{{\tt 1105.4749}}].

\bibitem{Luscher:2012av}
M.~{L\"uscher} and S.~Schaefer, \emph{{Lattice QCD with open boundary
  conditions and twisted-mass reweighting}},
  \href{http://dx.doi.org/10.1016/j.cpc.2012.10.003}{\emph{Comput.Phys.Commun.}
  {\bf 184} (2013) 519--528}, [\href{http://arxiv.org/abs/1206.2809}{{\tt
  1206.2809}}].

\bibitem{Bulava:2015bxa}
{\scshape ALPHA} collaboration, J.~Bulava, M.~Della~Morte, J.~Heitger and
  C.~Wittemeier, \emph{{Non-perturbative improvement of the axial current in
  $N_f$=3 lattice QCD with Wilson fermions and tree-level improved gauge
  action}},
  \href{http://dx.doi.org/10.1016/j.nuclphysb.2015.05.003}{\emph{Nucl. Phys.}
  {\bf B896} (2015) 555--568}, [\href{http://arxiv.org/abs/1502.04999}{{\tt
  1502.04999}}].

\bibitem{Bruno:2014lra}
M.~Bruno, P.~Korcyl, T.~Korzec, S.~Lottini and S.~Schaefer, \emph{{On the
  extraction of spectral quantities with open boundary conditions}},
  {\emph{PoS} {\bf LATTICE2014} (2014) 089},
  [\href{http://arxiv.org/abs/1411.5207}{{\tt 1411.5207}}].

\bibitem{Fritzsch:2012wq}
{\scshape ALPHA} collaboration, P.~Fritzsch, F.~Knechtli, B.~Leder,
  M.~Marinkovic, S.~Schaefer et~al., \emph{{The strange quark mass and Lambda
  parameter of two flavor QCD}},
  \href{http://dx.doi.org/10.1016/j.nuclphysb.2012.07.026}{\emph{Nucl.Phys.}
  {\bf B865} (2012) 397--429}, [\href{http://arxiv.org/abs/1205.5380}{{\tt
  1205.5380}}].

\bibitem{Schaefer:2010hu}
{\scshape ALPHA} collaboration, S.~Schaefer, R.~Sommer and F.~Virotta,
  \emph{{Critical slowing down and error analysis in lattice QCD simulations}},
  \href{http://dx.doi.org/10.1016/j.nuclphysb.2010.11.020}{\emph{Nucl.Phys.}
  {\bf B845} (2011) 93--119}, [\href{http://arxiv.org/abs/1009.5228}{{\tt
  1009.5228}}].

\bibitem{Korcyl:2016ugy}
P.~Korcyl and G.~S. Bali, \emph{{Non-perturbative determination of improvement
  coefficients using coordinate space correlators in $N_f=2+1$ lattice QCD}},
  \href{http://arxiv.org/abs/1607.07090}{{\tt 1607.07090}}.

\bibitem{Taniguchi:1998pf}
Y.~Taniguchi and A.~Ukawa, \emph{{Perturbative calculation of improvement
  coefficients to O(g**2a) for bilinear quark operators in lattice QCD}},
  \href{http://dx.doi.org/10.1103/PhysRevD.58.114503}{\emph{Phys.Rev.} {\bf
  D58} (1998) 114503}, [\href{http://arxiv.org/abs/hep-lat/9806015}{{\tt
  hep-lat/9806015}}].

\bibitem{Bulava:2016ktf}
J.~Bulava, M.~Della~Morte, J.~Heitger and C.~Wittemeier, \emph{{Nonperturbative
  renormalization of the axial current in $N_f = 3$ lattice QCD with Wilson
  fermions and a tree-level improved gauge action}},
  \href{http://dx.doi.org/10.1103/PhysRevD.93.114513}{\emph{Phys. Rev.} {\bf
  D93} (2016) 114513}, [\href{http://arxiv.org/abs/1604.05827}{{\tt
  1604.05827}}].

\bibitem{mattiatomasz}
M.~Dalla~Brida, T.~Korzec, S.~Sint and P.~Vilaseca, \emph{{High precision
  renormalization of the non-singlet axial current in lattice QCD with Wilson
  quarks}}, {\emph{{\emph{in preparation}}} }.

\bibitem{Campos:2015fka}
I.~Campos, P.~Fritzsch, C.~Pena, D.~Preti, A.~Ramos and A.~Vladikas,
  \emph{{Prospects and status of quark mass renormalization in three-flavour
  QCD}}, {\emph{PoS} {\bf LATTICE2015} (2016) 249},
  [\href{http://arxiv.org/abs/1508.06939}{{\tt 1508.06939}}].

\bibitem{Hasenfratz:2008fg}
A.~Hasenfratz, R.~Hoffmann and S.~Schaefer, \emph{{Reweighting towards the
  chiral limit}},
  \href{http://dx.doi.org/10.1103/PhysRevD.78.014515}{\emph{Phys.Rev.} {\bf
  D78} (2008) 014515}, [\href{http://arxiv.org/abs/0805.2369}{{\tt
  0805.2369}}].

\bibitem{Finkenrath:2013soa}
J.~Finkenrath, F.~Knechtli and B.~Leder, \emph{{One flavor mass reweighting in
  lattice QCD}}, \href{http://dx.doi.org/10.1016/j.nuclphysb.2013.10.019,
  10.1016/j.nuclphysb.2014.01.019}{\emph{Nucl.Phys.} {\bf B877} (2013)
  441--456}, [\href{http://arxiv.org/abs/1306.3962}{{\tt 1306.3962}}].

\bibitem{Bietenholz:2011qq}
W.~Bietenholz et~al., \emph{{Flavour blindness and patterns of flavour symmetry
  breaking in lattice simulations of up, down and strange quarks}},
  \href{http://dx.doi.org/10.1103/PhysRevD.84.054509}{\emph{Phys. Rev.} {\bf
  D84} (2011) 054509}, [\href{http://arxiv.org/abs/1102.5300}{{\tt
  1102.5300}}].

\bibitem{Bar:2013ora}
O.~{B\"ar} and M.~Golterman, \emph{{Chiral perturbation theory for gradient
  flow observables}}, \href{http://dx.doi.org/10.1103/PhysRevD.89.099905,
  10.1103/PhysRevD.89.034505}{\emph{Phys. Rev.} {\bf D89} (2014) 034505},
  Erratum: [Phys.\ Rev.\ D {\bf 89} (2014) no.9,  099905], 
  [\href{http://arxiv.org/abs/1312.4999}{{\tt 1312.4999}}].

\bibitem{Borsanyi:2012zs}
S.~Borsanyi et~al., \emph{{High-precision scale setting in lattice QCD}},
  \href{http://dx.doi.org/10.1007/JHEP09(2012)010}{\emph{JHEP} {\bf 1209}
  (2012) 010}, [\href{http://arxiv.org/abs/1203.4469}{{\tt 1203.4469}}].

\bibitem{Blum:2014tka}
{\scshape RBC, UKQCD} collaboration, T.~Blum et~al., \emph{{Domain wall QCD
  with physical quark masses}},
  \href{http://dx.doi.org/10.1103/PhysRevD.93.074505}{\emph{Phys. Rev.} {\bf
  D93} (2016) 074505}, [\href{http://arxiv.org/abs/1411.7017}{{\tt
  1411.7017}}].

\bibitem{Bazavov:2015yea}
{\scshape MILC} collaboration, A.~Bazavov et~al., \emph{{Gradient flow and
  scale setting on MILC HISQ ensembles}},
  \href{http://dx.doi.org/10.1103/PhysRevD.93.094510}{\emph{Phys. Rev.} {\bf
  D93} (2016) 094510}, [\href{http://arxiv.org/abs/1503.02769}{{\tt
  1503.02769}}].

\bibitem{Dowdall:2013rya}
R.~Dowdall, C.~Davies, G.~Lepage and C.~McNeile, \emph{{$V_{us}$ from $\pi$ and
  $K$ decay constants in full lattice QCD with physical u, d, s and c quarks}},
  \href{http://dx.doi.org/10.1103/PhysRevD.88.074504}{\emph{Phys.Rev.} {\bf
  D88} (2013) 074504}, [\href{http://arxiv.org/abs/1303.1670}{{\tt
  1303.1670}}].

\bibitem{Bruno:2013gha}
M.~Bruno and R.~Sommer, \emph{{On the $N_f$-dependence of gluonic
  observables}}, {\emph{PoS} {\bf LATTICE2013} (2013) 321},
  [\href{http://arxiv.org/abs/1311.5585}{{\tt 1311.5585}}].

\end{thebibliography}\endgroup
\end{document}